\documentclass[acmsmall]{acmart}

\setcopyright{cc}
\setcctype{by}
\acmJournal{PACMSE}
\acmYear{2026} \acmVolume{3} \acmNumber{ISSTA} \acmArticle{ISSTA073}
\acmMonth{10} \acmDOI{10.1145/3832164}

\usepackage{enumitem}
\usepackage{multirow}
\usepackage{xspace}
\usepackage{tcolorbox}

\usepackage{amssymb}
\usepackage{pifont}
\usepackage{breakurl}

\usepackage{url}
\usepackage{subcaption} %
\usepackage{wrapfig} %

\usepackage{makecell}

\newcommand{\tool}{\textsc{TDiff}\xspace}

\newcommand{\F}{Fig.}

\newcommand{\T}{Table}
\renewcommand{\S}{Sec.}

\usepackage{tablefootnote}

\usepackage[normalem]{ulem}

\newcommand{\mysubref}[2]{\hyperref[#1]{\ref*{#1}(#2)}}

\usepackage{tikz}
\newcommand* \circled [1] {\tikz [baseline=(char.base)]{
    \node[shape=circle, draw, inner sep=0.8pt] (char) {#1};}}

\newlist{researchquestions}{enumerate}{1}
\setlist[researchquestions]{label=\textbf{RQ\arabic*.},leftmargin=*}

\newcommand{\parh}[1]{\noindent\textbf{#1}}
\newcommand{\parhs}[1]{\noindent\underline{\textit{#1}}}

\begin{document}

\title{The Unseen Delta: Characterizing the Compiler Optimization Landscape via Top-Down Differential Analysis}

\author{Zhibo Liu}
\affiliation{%
  \institution{State Key Laboratory of Novel Software Technology, Nanjing University}
  \city{Nanjing}
  \country{China}}
\email{zhiboliu@nju.edu.cn}
\orcid{0000-0002-7872-1129}
\authornote{Corresponding author.}

\author{Huaijin Wang}
\affiliation{%
 \institution{Shandong University}
 \city{Qingdao}
 \country{China}}
\email{huaijinwang@sdu.edu.cn}
\orcid{0000-0002-1066-0331}

\author{Shuai Wang}
\affiliation{%
  \institution{The Hong Kong University of Science and Technology}
  \city{Hong Kong}
  \country{China}}
\email{shuaiw@cse.ust.hk}
\orcid{0000-0002-0866-0308}

\renewcommand{\shortauthors}{Liu et al.}

\begin{abstract}

  Compiler optimizations are essential for achieving high performance in modern
  software. However, recent studies highlight the persistence of performance
  bugs, i.e., subtle defects where the compiler generates functionally correct
  but computationally inefficient code, leading to significant performance
  degradation.
  Existing detection and testing methods typically employ a bottom-up approach,
  focusing on specific low-level code properties and remaining confined to known
  optimization rules. Consequently, they struggle to quantify the holistic %
  impact of identified issues and often overlook critical microarchitectural
  inefficiencies. %

  We observe a key indicator of untapped potential: different compilers often
  produce binaries with significant performance differences for identical source
  code. However, the root causes of these discrepancies remain largely
  unexplored and difficult to pinpoint using current techniques.
  To bridge this gap, we introduce a top-down differential analysis methodology.
  This approach calibrates compiler optimization differences with fine-grained,
  hierarchical microarchitectural metrics, offering a comprehensive view of
  runtime behavior.
  Using a sampling-based approach, this method efficiently pinpoints the
  critical code snippets responsible for performance differences, enabling
  targeted root cause analysis.

  Our empirical evaluation uncovers substantial and often surprising performance
  differences between binaries generated by GCC and Clang. A categorization of
  root causes reveals systemic challenges in compiler optimizations. %
  To quantitatively validate our findings and demonstrate practical impact, we
  developed a binary patching framework that fixes identified performance issues
  by transplanting superior code sequences from competing compilers.
  This work provides a novel lens for understanding and analyzing optimization
  defects.

\end{abstract}



\received{30 January 2026}
\received[accepted]{16 April 2026}

\maketitle %

\section{Introduction}
\label{sec:intro}

Modern compilers are cornerstone components of software engineering, %
translating high-level %
code into efficient, machine-executable instructions. To meet the performance
demands of contemporary computing, compilers employ a vast array of
sophisticated optimizations, encompassing techniques such as function
inlining~\cite{theodoridis2022understanding,chakrabarti2006inline,zhao2003inline,sewe2011next},
loop unrolling~\cite{rocha2020vectorization,huang1999generalized}, complex
instruction
scheduling~\cite{lozano2019survey,yoaz1999speculation,lozano2019combinatorial},
and automatic
vectorization~\cite{allen1988compiling,maleki2011evaluation,haj2020neurovectorizer,tian2017llvm}.
Given their critical role, the reliability of compiler optimizations has been
the subject of extensive
research~\cite{le2014compiler,le2015randomized,sun2016finding,chen2017learning,livinskii2020random,donaldson2021test,chen2022boosting,zhong2022enriching,livinskii2023fuzzing}.
Methodologies like differential and metamorphic testing have proven
pivotal in ensuring compilation correctness.
These efforts have successfully uncovered thousands of correctness
bugs~\cite{sun2016toward,chen2016empirical,chen2020survey,zhou2021empirical},
making modern compilers remarkably more robust.

However, beyond functional correctness lies the complex and less-explored domain
of performance optimization. Compilers may generate functionally correct code
that nevertheless exhibits substantial performance degradation, a phenomenon
attributed to compiler optimization defects. Unlike crashes or output errors,
these defects manifest as missed optimization opportunities or flawed heuristic
decisions, leading to suboptimal runtime performance, increased memory
consumption, or excessive binary size. 
Prior research has sought to identify such
issues~\cite{barany2018finding,gong2018empirical,tan2020every,theodoridis2022finding,theodoridis2024refined,gao2024shoot},
primarily through ``bottom-up'' methods that scrutinize source code or binaries
for predefined, specific optimization anti-patterns. Such approaches target
concrete instances where, for example, a loop was not
vectorized~\cite{gong2018empirical}, or a function was not inlined despite
apparent suitability~\cite{theodoridis2022understanding}.

While valuable for detecting specific, localized flaws, bottom-up techniques
are inherently constrained. They can identify a missed optimization,
but cannot quantify its holistic impact on end-to-end program
performance. 
A key reason is that modern CPU microarchitectures can often mitigate
low-level code inefficiencies through hardware mechanisms such as pipelining and
out-of-order execution.

\begin{figure*}[h]
    \centering
    \includegraphics[width=0.999\linewidth]{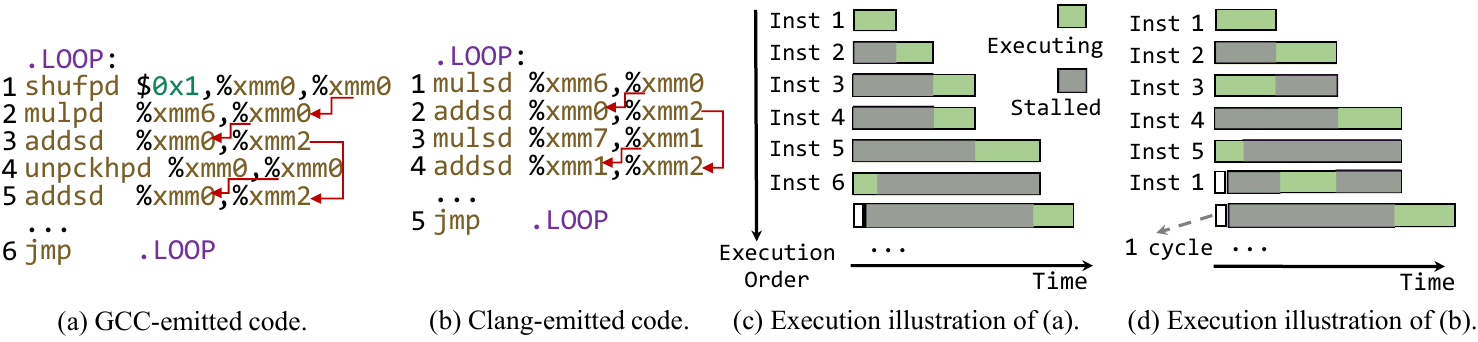}

    \caption{A simplified illustration of CPU pipelining's impact on
    performance. Despite code (a) having more instructions and a longer
    dependency chain than code (b), %
    their throughput is close. %
    Due to superscalar out-of-order execution, the time
    between two iterations is 1 cycle for both, resulting in equal overall
    performance.}
    \label{fig:intro}
\end{figure*}

\F~\ref{fig:intro} illustrates how these hardware features can mask performance
differences that appear significant at the instruction level. The figure
compares two semantically equivalent code snippets within a loop, one generated by
GCC (\F~\mysubref{fig:intro}{a}) and one by Clang (\F~\mysubref{fig:intro}{b}). 
An analysis focusing on low-level code metrics (e.g., instruction
count) would favor \F~\mysubref{fig:intro}{b}, as it contains one fewer instruction
(5 vs. 6), suggesting a 16\% performance advantage.
However, modern superscalar processors execute independent instructions in
parallel, dynamically reordering them to maximize hardware utilization.
As shown in the execution diagrams (\F~\mysubref{fig:intro}{c} and
\mysubref{fig:intro}{d}), although the data dependency chains differ, the processor
pipeline allows a new loop iteration to begin 
every cycle. Consequently, despite differences in instruction counts, the
throughput of both loops is close, resulting in identical overall runtime
performance.

This discrepancy between low-level code properties and runtime execution
behavior obscures critical performance realities and underscores the limitations
of %
bottom-up approaches. 
Although modern compilers are known
to generate distinct code patterns for the same semantics, the precise mapping
between these low-level variances and actual runtime performance remains a
complex and open challenge.
Our work is therefore motivated by the need for a
holistic, comprehensive performance analysis methodology and is structured around the
following research questions: %

\begin{itemize}[noitemsep,leftmargin=3mm] %
	\item \textbf{RQ1:} To what extent do optimization defects
    identified by existing bottom-up methods affect overall program runtime
    performance?
	
    \item \textbf{RQ2:} To what extent do optimization differences between
    mature, state-of-the-art compilers yield significant performance
    disparities?
	
    \item \textbf{RQ3:} How can we effectively identify critical code snippets
    responsible for significant performance differences to diagnose their root
    causes?
\end{itemize}

Answering these questions requires a paradigm shift. Our intuition is that
existing approaches are inadequate for three reasons. First, coarse-grained
code-level metrics 
are poor proxies for runtime performance, as they fail to capture the complex
interplay between code and the underlying hardware microarchitecture.
Second, by focusing on predefined patterns, bottom-up methods preclude the
discovery of novel or emergent sources of performance loss. 
Finally, methods based on source code mutation or synthetic benchmarks (e.g.,
programs randomly generated with CSmith~\cite{yang2011finding} or
YARPGen~\cite{livinskii2020random}) often fail to detect defects that manifest
in large, real-world programs and do not provide a clear pathway for actionable
remediation. 
To comprehensively characterize optimization defects, 
we must (1) prioritize overall, end-to-end performance as
the primary signal, (2) monitor program execution with fine-grained, hierarchical
microarchitectural metrics, and (3) precisely pinpoint code
locations that contribute most to performance degradation.

To address these limitations, we introduce a top-down differential
analysis methodology designed to diagnose inefficiencies caused by optimization
defects. This process begins by identifying significant runtime variations
between binaries generated by modern compilers (e.g., GCC vs. Clang).
It then progressively refines the analysis by coalescing diverse hierarchical
microarchitecture events, such as cache misses, branch mispredictions, and
pipeline stalls, collected via
hardware performance monitors~\cite{guide2011intel,yasin2014top}. 
This approach enables a multidimensional performance analysis that reveals
novel insights that are often obscured at the source or binary levels. By
leveraging hardware-assisted
sampling~\cite{sasongko2023precise}, this method effectively isolates critical
code snippets responsible for performance disparities, facilitating both deep
root-cause analysis and practical remediation.

Through an extensive manual investigation of pinpointed suboptimal code
snippets, we identify and classify 
eight categories of low-level code patterns
that repeatedly cause performance divergence. We provide detailed discussions
with concrete code examples to characterize these systemic issues in compiler
optimizations.
Our findings demonstrate that cross-compiler analysis remains a potent avenue
for diagnosing and improving optimization strategies.

In summary, this paper makes the following contributions:
\begin{itemize}[noitemsep,topsep=0pt,leftmargin=3mm] 

    \item Conceptually, we introduce and advocate for a top-down, impact-driven
    paradigm for studying compiler optimization defects, shifting the focus from
    isolated anti-patterns to holistic, microarchitecture-aware performance
    analysis.

    \item Technically, we design and implement \tool, a \textbf{t}op-down
    \textbf{diff}erential analysis framework that leverages hardware performance
    counters and sampling to pinpoint critical code snippets responsible for
    performance gaps effectively. We further develop a binary patching pipeline
    to validate our findings and demonstrate practical applicability.

    \item Empirically, we conduct an in-depth study using real-world benchmarks,
    revealing substantial performance differences ranging from 5\% to
    80.8\% between code generated by the latest versions of GCC
    (15.1.0~\footnote{Released on April 4, 2025}) and Clang
    (20.1.0~\footnote{Released on March 5, 2025}). We provide a taxonomy
    of root causes for these differences and
    quantitatively demonstrate the practical impact via our binary patching
    framework.

\end{itemize}

\section{Background}
\label{sec:bg}

\subsection{Compiler Testing and Optimization Defects}

To contextualize this study's contributions, we first review established
methodologies for analyzing compiler optimization defects.
Recent research increasingly focuses on analyzing subtle differences in compiler
optimizations across multiple dimensions. 
Notably, Theodoridis et
al.~\cite{theodoridis2022finding,theodoridis2022understanding,theodoridis2024refined}
identified significant optimization differences and defects in modern compilers
that substantially impact binary size and performance. Their findings reveal
widespread optimization flaws and highlight a %
vast, under-explored research area.

Inspired by recent works, we explore new paradigms for deeply investigating
defects in modern compiler optimizations and uncovering potential optimization
opportunities.
Prevailing techniques predominantly employ bottom-up analysis but vary
significantly in their targets and metrics. As summarized in \T~\ref{tab:bg}, we
categorize them broadly into binary-level and source-level methods.

\begin{table}[!htp]
	\centering
	\caption{A comparison with recent works.} 

    \resizebox{0.67\linewidth}{!}{
		\begin{tabular}{l c  c  c }
			\hline 
			& Recent Works & Target Programs  & Research Focuses  \\
			\hline
            \multirow{3}{*}{Binary} & Barany et al.~\cite{barany2018finding} & \multicolumn{1}{|c|}{synthetic}  & custom binary features  \\
            & Theodoridis et al.~\cite{theodoridis2022finding} & \multicolumn{1}{|c|}{synthetic}  & dead code elimination  \\
            & Theodoridis et al.~\cite{theodoridis2022understanding} & \multicolumn{1}{|c|}{real-world}  & function inlining  \\
            
            \hline
			\multirow{4}{*}{Source} & Gong et al.~\cite{gong2018empirical} & \multicolumn{1}{|c|}{loop nests}  & loop transformations  \\	
            & LoopLearner~\cite{mammadli2021learning} & \multicolumn{1}{|c|}{loop nests}  & loop transformations  \\
			& Theodoridis et al.~\cite{theodoridis2024refined} & \multicolumn{1}{|c|}{synthetic}  & source info injecting \\
            & de3~\cite{gao2024shoot} & \multicolumn{1}{|c|}{real-world}  & source transformations  \\
            \hline
			Our & \tool & \multicolumn{1}{|c|}{real-world}  & microarchitectural behaviors \\
			\hline
		\end{tabular}
	}
	\label{tab:bg}

\end{table}

\noindent \textbf{Binary-level approaches} analyze generated
machine code using specific performance indicators. 
A prevalent strategy is differential testing, which compares outputs across
varying compilers or optimization levels.
For instance, Barany et al.~\cite{barany2018finding} pioneered
instruction-level differential testing to identify missed
optimizations through cross-compiler comparisons. 
More recently, Theodoridis et al.~\cite{theodoridis2022finding} utilized dead
code elimination (DCE) as a lens for differential testing; specifically, if one
compiler retains a dead code block that another compiler eliminates, it
indicates a potential missed optimization.
Other research examines high-impact optimizations to
expose the limitations of compiler heuristics.
Theodoridis et al.~\cite{theodoridis2022understanding} developed
an exhaustive search method for optimal function inlining decisions with a focus
on reducing binary size. 
While effective at detecting localized flaws, these binary-centric methods are
often constrained by their reliance on predefined code properties and fail
to capture holistic performance implications.

\noindent \textbf{Source-level approaches} initiate analysis by transforming
input source code.
These methods create multiple semantically equivalent program variants and
compare the performance of compiled binaries. The underlying
premise is that a robust compiler should generate consistently efficient code
despite superficial syntactic changes. 
Gong et al.~\cite{gong2018empirical} empirically evaluated compiler stability by
applying diverse source transformations to extracted loop nests and
measuring performance disparities. 
Similarly, Gao et al.~\cite{gao2024shoot} investigated how seemingly efficient
source code can inadvertently induce compiler inefficiencies, %
employing source transformations as an investigative mechanism. 
In related work, Theodoridis et al.~\cite{theodoridis2024refined}
demonstrated that providing compilers with additional information (e.g.,
explicitly marking unreachable paths) can paradoxically degrade performance
due to unforeseen interactions between optimization phases.
To navigate the vast search space of transformations, Mammadli et
al.~\cite{mammadli2021learning} introduced a neural network-based approach that
learns to predict source-level loop transformations to facilitate the generation
of more efficient code.

\parh{Limitations.}
Despite uncovering pervasive optimization defects, both source- and binary-level
bottom-up methods share critical limitations.
As noted in \S~\ref{sec:intro}, static code features and narrowly defined
anti-patterns do not consistently correlate with real-world performance gaps. 
Furthermore, reliance on predefined code properties or transformations
constrains the discovery of novel defect categories. 
Critically, most existing methodologies rely on extracted loops or synthetic
programs, making them difficult to apply to complex real-world software.
Although some studies target real-world
applications~\cite{gao2024shoot,theodoridis2022understanding}, they are
restricted to specific optimization categories (e.g., function inlining) or
limited to manually defined code transformations. 
To address these limitations, we introduce a holistic, end-to-end analysis
method that examines microarchitectural behaviors in depth, enabling broad
coverage of diverse code properties with potential performance implications.

\subsection{Hardware Performance Monitoring}

The gap between static code properties and actual runtime behavior stems
from the complexity of modern CPU microarchitectures. Features like deep
instruction
pipelines~\cite{hartstein2002optimum,sprangle2002increasing,finlayson2013improving},
out-of-order execution~\cite{golden201140,padmanabha2017mirage}, speculative
execution~\cite{gabbay1996speculative,maisuradze2018ret2spec,kocher2020spectre},
and multi-level memory
caches~\cite{baer1988inclusion,przybylski1989characteristics,nori2018criticality}
make accurate performance prediction by merely inspecting code alone
difficult.

To provide deep visibility into subtle microarchitectural operations,
modern CPUs integrate performance monitoring units
(PMUs)~\cite{contreras2005power,guide2011intel}, also known as 
hardware performance counters~\cite{weaver2013non,das2019sok}. 
PMUs are special-purpose registers that count a wide range of low-level hardware
events, such as different levels of cache misses, branch mispredictions,
pipeline stalls, and floating-point operations, with minimal overhead. 
By providing direct measurements of how efficiently the hardware is being
utilized, these counters facilitate substantially deeper and more precise
assessment of program performance than simple wall-clock time. 

\parh{The Top-Down Hierarchy.}
Building on this hardware-monitoring capability, 
methodologies such as 
the top-down microarchitecture analysis method
(TMAM)~\cite{yasin2014top,jarus2016top,mou2024top} have emerged to diagnose
performance bottlenecks systematically. Originally proposed by Yasin
for Intel architectures~\cite{yasin2014top},
TMAM organizes the vast number of raw hardware events into a hierarchical
decision tree, allowing analysts to navigate from high-level execution overviews 
progressively down to specific root causes.
By adopting this structured, top-down paradigm, we shift our analytical focus
from syntactic code characteristics to observed hardware behaviors, %
establishing the %
foundation for the microarchitecture-aware optimization defect analysis proposed
in this work.

\section{The Predicament of Optimization Defect Detection}
\label{sec:motivation}

Effective tests for compiler optimization defects should prioritize finding
defects that significantly impact program performance and are representative of
common real-world issues. Identifying these defects allows developers to fix
critical performance bugs and improve compiler optimizations. 
The core challenge, however, lies in systematically uncovering these subtle
flaws and accurately quantifying their impact.
As discussed in \S~\ref{sec:intro}, while valuable, prevailing bottom-up
approaches may fail to accurately capture the true real-world impact of
optimization defects.

\subsection{Deviation from Impactful Optimization Defects}

To justify the motivation for a new top-down testing paradigm, this
section first presents a preliminary study %
to understand the gaps between prevailing
testing methodologies and the detection of impactful optimization defects.
Specifically, we investigate RQ1 by collecting and analyzing the optimization
defects uncovered in prior
works~\cite{barany2018finding,theodoridis2022finding,gao2024shoot,theodoridis2024refined}.\footnote{While
some prior works also evaluate compiled binary size, our analysis focuses solely
on runtime performance. 
We make this distinct to contextualize our contributions
rather than to underestimate the significance of prior research.
Nonetheless, the performance defects we uncover provide a compelling motivation
for the proposed new paradigm.}
Nevertheless, reproducing these defects is challenging. Many of them are
dependent on specific compiler versions, target architectures, and compilation
flags, making successful recompilation non-trivial. 
Moreover, the performance impact of these defects is often difficult to measure,
as reported programs are often largely reduced and lack the necessary execution
context or input data.

To facilitate manual analysis, we restricted our scope to reported optimization
defects that include both compilable source code and the corresponding assembly
code. To evaluate the performance impact of these defects in an
environment-agnostic manner, we leveraged uiCA~\cite{abel2022uica}, a
simulation-based throughput predictor. uiCA models the characteristics of the
CPU pipeline and out-of-order execution to statically estimate the average CPU
cycles per machine instructions.\footnote{As uiCA does not currently support
ARM, we also utilized \texttt{llvm-mca}~\cite{llvm-mca} to analyze ARM
assembly.} 
uiCA estimates steady-state instruction throughput under an idealized
memory system, and it does not model dynamic microarchitectural events such as
cache or TLB misses and branch mispredictions. Its estimates should therefore be
interpreted as a throughput-oriented lower bound rather than a measurement of
end-to-end runtime impact. Given these limitations, we relied primarily on
manual analysis to draw conclusions, using the simulated results only as
supplementary evidence. A full assessment on real hardware is infeasible, as the
reported defects are heavily reduced and lack execution drivers or inputs.

\begin{figure}[h]
    \centering
    \includegraphics[width=0.999\linewidth]{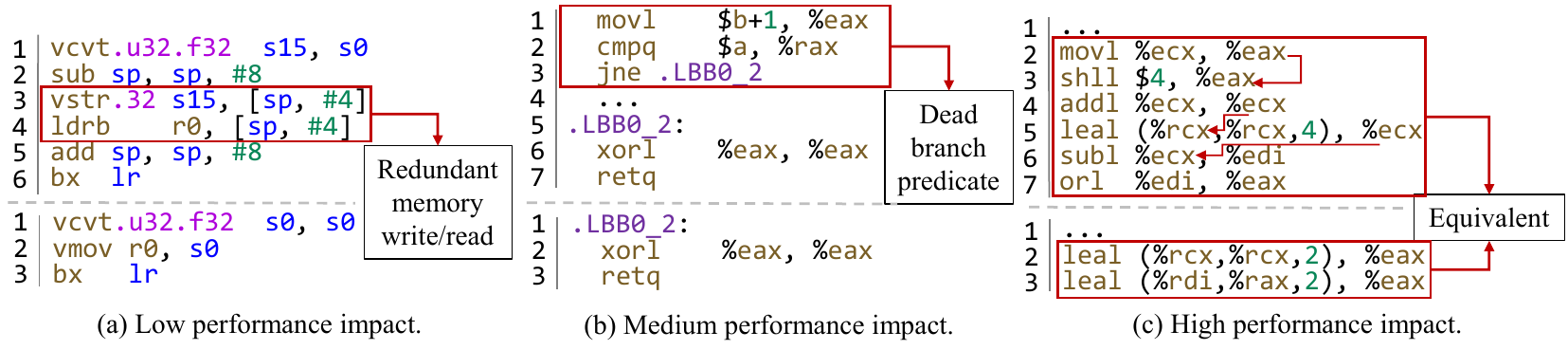}

    \caption{Illustrative examples of optimization defects with different impact
	levels. The suboptimal code is shown above the optimized version.}
    \label{fig:motivate}

\end{figure}

Overall, we collected 36 optimization defects with reported code cases
identified in prior literature.
Using uiCA, we analyze differences in CPU cycles by simulating 100
iterations of the corresponding assembly snippets. This simulates the cumulative
performance overhead  
of frequently executed (``hot'') code, mirroring performance bottlenecks found
in real-world scenarios.
In addition, since these code snippets often lack execution context, we classify
them into three categories (as shown in \F~\ref{fig:motivate}) to better assess
their potential impact on overall program performance.

We divide these defects into three tiers based on their impact on runtime
performance. First, as illustrated in \F~\mysubref{fig:motivate}{a}, low-impact
defects involve inefficiencies such as redundant memory accesses. The latency of
these operations is often masked by the CPU cache and pipeline, resulting in
negligible overhead. In this example, although the suboptimal code
contains a higher instruction count and two redundant memory accesses, the data
is primarily served from cache during execution. Furthermore, because the code
exhibits low execution port pressure and lacks loop-carried data dependencies,
the processor can keep the execution units fully saturated. Consequently, it
achieves throughput comparable to the optimized version.
Second, the medium-impact category encompasses inefficiencies that introduce a
noticeable, constant overhead that scales linearly with the number of loop
iterations. For example, in \F~\mysubref{fig:motivate}{b}, an unoptimized dead
branch predicate imposes a fixed cost of three instructions per iteration, which
is amortized over code in the loop.
Finally, the high-impact category comprises defects such as suboptimal
arithmetic computations (\F~\mysubref{fig:motivate}{c}). Such defects not only
incur significant overhead but also frequently appear in computationally
intensive kernels, implying a substantial real-world impact.

\T~\ref{tab:motivation-defects} summarizes the results. After excluding
two cases incompatible with uiCa and \texttt{llvm-mca}, 
we find that only 11 out of 34 cases (32.4\%) have a high performance impact.
Notably, these high-impact defects are often related to suboptimal arithmetic
computations and specific code patterns (e.g., missed vectorization,
redundant initialization, missed function inlining, and unrecognized patterns of
standard library functions like \texttt{memset}).
Separating the results by program source reveals a sharp disparity:
85.7\% (6/7) of the defects reported from real-world programs are high-impact,
compared with only 18.5\% (5/27) of those from synthetic programs.

\smallskip
\begin{tcolorbox}[size=small,colback=gray!5!white,colframe=gray!75!black,]
\parh{Finding 1 (RQ1):}~
Under a controlled throughput model, only 32.4\% of the 
defects exhibit a high impact.
The results indicate that code-level performance
indicators crafted by %
bottom-up methods may fail to capture the most impactful defects,
potentially steering developers towards relatively low-impact issues.
This shortcoming is exacerbated by the reliance on synthetic benchmarks, which
can be ill-suited to revealing impactful defects in %
real-world applications.
\end{tcolorbox}

\begin{table}[!htp]
	\centering
	\caption{Performance impact assessment of reported optimization defects.}
	\resizebox{0.649\linewidth}{!}{
	\begin{tabular}{l r | c c c c | c c}
    \hline
    \multirow{3}{*}{Impact} & \multirow{3}{*}{\#Cases}
      & \multicolumn{4}{c|}{Root-cause characteristic}
      & \multicolumn{2}{c}{Program source} \\
    \cline{3-6} \cline{7-8}
      & & \makecell{Mem.\\Access} & \makecell{Arith.\\Comp.}
        & \makecell{Branch\\Pred.} & Others
        & Synthetic & Real-world \\
    \hline
    Low    & 5  & 5 & 0 & 0  & 0 & 4  & 1 \\   
    Medium & 18 & 0 & 2 & 16 & 0 & 18 & 0 \\   
    High   & 11 & 0 & 6 & 0  & 5 & 5  & 6 \\
    \hline
    Total  & 34 & 5 & 8 & 16 & 5 & 27 & 7 \\
    \hline
    \end{tabular}
	}
	\label{tab:motivation-defects}
\end{table}

\subsection{Widespread Performance Divergence} %
\label{sec:motivation-compiler}

While analyzing known defect patterns offers limited insight into overall
performance, a comparison of modern compilers reveals more substantial
divergences.
To investigate RQ2, we compiled a suite of real-world benchmarks, including
Polybench~\cite{pouchet2012polyhedral},
Coremark-Pro~\cite{gal2012exploring,coremark-pro}, the C benchmarks
from SPEC CPU 2017~\cite{spec2017}, and a collection of high-performance
computing (HPC) programs (detailed in \S~\ref{sec:experiments-setup}),
using the latest GCC and Clang at the \texttt{-O3} optimization level. 

We meticulously measure wall-clock execution time of binaries  
while minimizing noise from the operating system, other applications, and
hardware by applying several techniques. 
This includes disabling address space layout randomization (ASLR), running
benchmarks in single-thread mode, and attaching programs to a fixed CPU core to
reduce CPU and scheduler noise. 
We run each program 10 times and record
the time of the fastest run to mitigate the potential impact of cache
conflicts due to background noise.
We also focus only on performance disparities exceeding 5\%.

\begin{table}[!htp]
		\centering
		\caption{Runtime performance differences between binaries generated by different compilers.} 

		\resizebox{0.507\linewidth}{!}{
		\begin{tabular}{l | c | c  c  c  c}
			\hline 
			\multirow{2}{*}{Benchmarks} & \multirow{2}{*}{\#Prog.} & \multicolumn{4}{c}{\#Prog. with performance difference} \\
			\cline{3-6}
			& & 5\%-10\% & 10\%-30\% & over 30\% & sum \\
			\hline
			Polybench    & 30  & 4 & 3 & 0  & 7 \\
			Coremark-Pro & 9 & 4 & 4 & 0 & 8 \\
			SPEC & 8 & 2 & 2 & 1  & 5 \\
			HPC Apps   & 7 & 1 & 3 & 3  & 7 \\
			\hline
		\end{tabular}
		}
        \label{tab:motivation-performance}
\end{table}

As shown in \T~\ref{tab:motivation-performance}, we found substantial and often
unpredictable performance differences between code generated by the two
compilers (detailed in \S~\ref{sec:experiments-overall}), with neither
being consistently superior across all benchmarks. 
Although the benchmarks we evaluated are commonly used in compiler research, we
still find that 27 out of 54 programs (50\%) exhibit a performance
difference greater than 5\%, and
16 (29.6\%) show a difference greater than 10\%.

\smallskip
\begin{tcolorbox}[size=small,colback=gray!5!white,colframe=gray!75!black,]
\parh{Finding 2 (RQ2):}~This widespread performance delta highlights a vast,
underexplored landscape for potential optimization. The fact that one compiler
can generate significantly faster code for the same source proves that a more
optimal solution exists. 
However, current techniques cannot systematically investigate
these differences. They fail to pinpoint the exact code locations 
responsible for the performance gap and, more importantly, 
the specific underlying microarchitectural reasons, 
leaving developers without a clear path to understand
and bridge these performance gaps. This ``unseen delta'' is the
primary motivation for our top-down differential analysis methodology.
\end{tcolorbox}

\section{Empirical Study of Cross-Compiler Performance Differences}
\label{sec:experiments}

\begin{figure*}[!htp]

    \centering
    \includegraphics[width=0.999\linewidth]{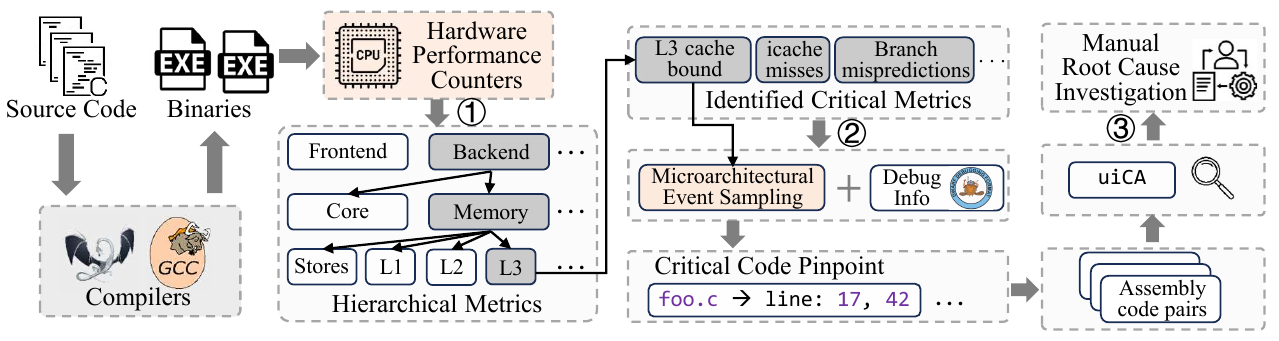}

    \caption{Top-down differential analysis pipeline. 
    We employ a top-down approach with a microarchitectural metric hierarchy to
    identify critical metrics that show significant differences between
    binaries. Based on these metrics, hardware event sampling then pinpoints the
    critical code snippets driving the behavioral divergence. Finally, root
    causes of the performance differences are diagnosed through
    uiCA-assisted manual analysis. }

    \label{fig:pipeline}
\end{figure*}

To thoroughly investigate compiler performance differences (RQ2) and
identify their root causes (RQ3), we propose a paradigm shift
from traditional bottom-up to top-down differential analysis. 
This approach prioritizes end-to-end performance impact, using hierarchical
microarchitectural metrics to trace optimization defects systematically. 
Unlike existing methods, our methodology offers a holistic, performance-centric
view of compiler optimizations. It complements prior works by tracing
high-level performance deltas down to the specific microarchitectural behaviors
and code snippets responsible. 
While orthogonal to source transformation
techniques~\cite{sun2016finding,li2024boosting,zhang2017skeletal,wu2025compiler,theodoridis2024refined,gao2024shoot}, this framework can integrate with them to
explore a broader optimization space.
This section details our methodology, experimental setup, and cross-compiler
comparison results.

\subsection{Top-Down Differential Analysis Methodology}
\label{sec:method}

Our analysis pipeline, illustrated in\F~\ref{fig:pipeline}, systematically
drills down from overall performance differences to specific root causes in
three stages. 
We begin with binaries compiled from identical source code by different
compilers. First, we identify ``critical metrics''---microarchitectural
performance metrics exhibiting a divergence greater than 5\%.
Using hardware-assisted event sampling, we then pinpoint critical code snippets
driving these differences. 
Finally, we manually investigate these snippets to diagnose the underlying root
causes of inefficiencies.

\parh{Hierarchical Performance Metric Analysis.}~Modern processors are equipped
with PMUs to track low-level events such as cache misses or branch mispredictions.
Given the sheer volume of available events~\cite{intel64ia32sdm}, we employ an
established hierarchical performance model~\cite{yasin2014top} to structure our
analysis and systematically pinpoint bottleneck microarchitectural behaviors (\circled{1}). 
Specifically, at the highest level, %
a CPU's execution pipeline slot falls into one of four states: frontend
bound, bad speculation, retiring (useful operation completion),
or backend bound. 
Comparing two binaries, a performance issue in one will manifest as a
disproportionate amount of time spent in one category, immediately narrowing the
root-cause search space.

\begin{wrapfigure}[17]{l}{0.61\linewidth}
    \centering
    \includegraphics[width=0.98\linewidth]{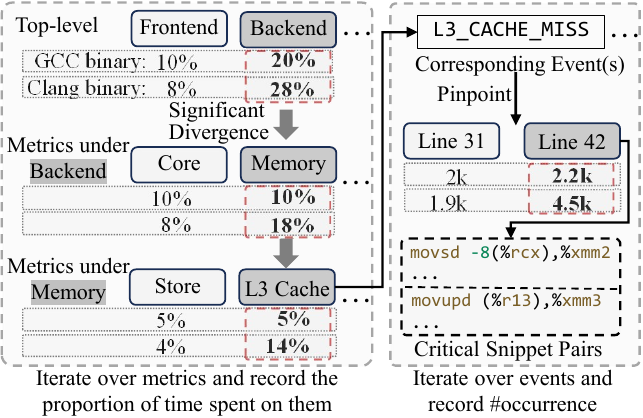}

    \caption{Illustrative example of the top-down process.}
    \label{fig:method-example}

\end{wrapfigure}

This hierarchy's strength lies in its drill-down capability. 
As illustrated in \F~\ref{fig:method-example},
once a
top-level category (e.g., backend bound) is identified, we descend to differentiate
its subcategories, for example, core bound (stalls from computation/resource contention) or
memory bound (stalls from memory access), by examining the proportion of each
subcategory.
A divergent proportion of time spent on memory-bound issues directs
the analysis further down the hierarchy to specific hardware behaviors, such as L3
cache misses.

This structured use of diverse metrics offers clear advantages. It creates a
repeatable, data-driven workflow that directs attention to primary performance
limiters, avoiding distraction by secondary issues. 
Insights gained are actionable, as each hierarchical node (i.e., a metric)
corresponds to specific hardware behaviors, suggesting targeted
remedies. 
Ultimately, the hierarchical model resolves ambiguities of isolated metrics,
offering a holistic view of how code-microarchitecture interactions lead to
performance degradation.

\parh{Sampling-Based Pinpointing.}
After identifying the divergent microarchitecture metrics, we use hardware-based
event sampling to locate the specific code fragments responsible for each
metric's deviation (\circled{2}).
This technique leverages PMUs for low-overhead profiling, constructing a
statistical profile of a program's execution and resource utilization patterns
without significantly impeding its performance.
Instead of continuously monitoring all microarchitectural events, which would
generate unmanageable event records, we configure the PMU to monitor a specific
event and trigger a hardware interrupt after a predefined number of occurrences.
At each interrupt, the operating system captures the execution context,
particularly the instruction pointer that indicates the instruction being
executed at the time of interrupt.

This method's key advantages are its generalizability across diverse
microarchitectural metrics and minimal performance impact. 
Microarchitectural behaviors are highly sensitive to low-level code patterns;
thus, common intrusive profiling
methods~\cite{luk2005pin,nethercote2007valgrind,duck2020binary} can distort
program behavior and yield misleading results.
In contrast, hardware-managed event counting with limited %
software intervention minimizes disturbance to the program's microarchitectural
behaviors, avoiding the ``observer effect'' and enabling practical profiling of
large, performance-sensitive applications. %

The sampling output comprises a collection of raw memory addresses marking
interrupt locations. Using compiler-generated debug information, we map these
addresses back to source code locations (e.g., source files and line numbers).
The debug information acts as a bidirectional link between
the low-level machine code generated by different compilers, by associating
ranges of machine instruction addresses with their corresponding source code
locations.

Pinpointing critical code locations involves aggregating collected samples.
The profiling typically results in the thousands or millions of captured
instruction pointers. By attributing these addresses to specific source
locations, we can effectively determine code snippets %
exhibiting significantly inconsistent behavior between binaries, enabling
targeted analysis of the root cause.
Specifically, as illustrated in \F~\ref{fig:method-example}, given a code
snippet located in one binary that frequently triggers a certain type of event,
if its counterpart in another binary triggers significantly fewer events, we
deem these snippets critical and responsible for the observed microarchitectural
behavior differences.

\parh{Microarchitecture-Aware Root Cause Investigation.}
The final stage is a deep dive into defects. Given a pair of critical code
snippets exhibiting performance disparities, we move beyond identifying where
the issue occurs to precisely understanding why specific code
patterns result in low efficiency. This involves examining the generated machine
code and modeling its interaction with CPU microarchitecture (\circled{3}).

We compare specific machine instructions, assess
instruction-level parallelism, and analyze memory access patterns that could
incur high latency.
We also evaluate the presence or absence of major
optimizations (such as vectorization) that dramatically impact performance.
To quantify these defects, we utilize a microarchitecture throughput predictor
(e.g., uiCA) on assembly code. %
This allows us to model instruction throughput, resource utilization, and data
dependencies, revealing subtle yet impactful differences in %
low-level code patterns that explain the performance gaps. %

However, throughput predictors are limited by their inability to
model performance penalties associated with dynamic runtime states.
Therefore, we note that \tool\ does not automatically identify root causes of
performance disparities. 
Instead, \tool\ isolates the problem area and provides reference
throughput data, facilitating a manual expert analysis.
This combination of automated profiling and expert interpretation is
indispensable, as the true cause of performance degradation often relies on
complex code-hardware interactions that require deep microarchitectural insight
to diagnose.
This detailed investigation not only diagnoses the characteristics of specific
defects but also forms the basis for categorizing root causes, as discussed in
\S~\ref{sec:analysis}.

\subsection{Experimental Setup}
\label{sec:experiments-setup}

To validate our top-down microarchitecture-aware differential analysis approach,
we extend the preliminary study from
\S~\ref{sec:motivation-compiler} with a comprehensive investigation. We conduct
an in-depth analysis to evaluate the effectiveness of our approach in
locating and diagnosing optimization defects.
Meanwhile, to ensure the reliability and validity of our findings, we follow the
setup described in \S~\ref{sec:motivation-compiler} to establish a controlled
experimental environment.

\parh{Benchmarks.}~We use a collection of 54 real-world benchmarks
widely adopted in compiler and high-performance computing research.
This suite includes 30 numerical computation kernels from
Polybench~\cite{pouchet2012polyhedral}, 9 applications from
Coremark-Pro~\cite{coremark-pro} (\T~\ref{tab:coremark}), a comprehensive
industry-standard processor benchmark, and all 8 pure C benchmarks from SPEC CPU
2017~\cite{spec2017}.
While these benchmarks represent a broad range of domains (e.g., data
compression, cryptography, and neural network evaluation), we further expanded
our dataset with 7 well-established HPC applications, listed in
\T~\ref{tab:hpc-description}. 
We selected these benchmarks because they contain performance-critical
computations that stress various aspects of compiler optimizations.

\parh{Compilation.}~We compiled all programs using GCC 15.1.0 and
Clang/LLVM 20.1.0 with the \texttt{-O3} flag. While modern compilers
offer processor-specific options such as \texttt{-mtune} and \texttt{-march},
our %
observations indicate that \texttt{-mtune} 
provides no discernible performance benefit compared with \texttt{-O3}. 
Conversely, while \texttt{-march} allows for significant gains via modern
instruction set extensions, it sacrifices backward compatibility, limiting its
adoption in prebuilt binaries.
Therefore, we focused on the ubiquitous \texttt{-O3} configuration to reflect
the most prevalent deployment scenarios. Nevertheless, 
our methodology
remains applicable to uncovering defects in processor-specific
optimizations.

\parh{Implementation.}~Our analysis framework, \tool, automates the end-to-end
workflow, including compilation, execution, data collection, and differential
analysis. \tool\ uses the standard Linux \texttt{perf} utility for
microarchitectural event monitoring and uiCA to facilitate manual
root cause analysis. 
Since tools like \texttt{perf-diff} are ill-suited for comparing
binaries generated by different compilers or detecting optimization defects, we
implemented a custom comparison mechanism based on debug information (as
discussed in \S~\ref{sec:method}).
While our evaluation uses an Intel Core i7-9700K with 32 GB RAM on Ubuntu 22.04
LTS, \tool's modular design supports easy integration of new hardware and
compilers.

\begin{table}[!htp]
    \centering

    \begin{minipage}{0.435\textwidth}
        \centering
        \caption{Descriptions of Coremark-Pro benchmarks.}
        \label{tab:coremark}

        \resizebox{0.99\linewidth}{!}{
        \begin{tabular}{c|l}
            \hline
            Name & Description   \\
            \hline
            \texttt{core} & A collection of common algorithms \\
            
            \texttt{cjpeg} & JPEG image compression  \\
            
            \texttt{sha256} & Secure hash algorithm  \\
            \texttt{linpack} & Gaussian elimination with partial pivoting  \\
            \texttt{nnet} & Neural net  \\
            \texttt{loops} & Kernels based on Livermore loops  \\
            \texttt{parser} & XML parsing  \\
            \texttt{radix2} & Radix-2 fast Fourier transform  \\
            \texttt{zlib} & ZIP compression  \\
            
            \hline
        \end{tabular}
        }
    \end{minipage}
    \hfill %
	\begin{minipage}{0.53\textwidth}
        \centering
        \caption{Descriptions of HPC benchmarks.}
        \label{tab:hpc-description}

        \resizebox{0.99\linewidth}{!}{
        \begin{tabular}{c|l}
            \hline
            Name & Description   \\
            \hline		
            \multirow{2}{*}{\texttt{lbm}~\cite{hpc1}} & Lattice Boltzmann method for computational  \\
                                                      & fluid dynamics \\
            
            \multirow{2}{*}{\texttt{flow}~\cite{hpc2}} & A 2D hydrodynamics miniapp using Lagrangian-   \\
                                                       & Eulerian method   \\
            \texttt{hot}~\cite{hpc3} & A heat diffusion miniapp with a CG solver  \\

            \multirow{2}{*}{\texttt{simpleMOC}~\cite{hpc4}} & A 3D method of characteristics (MOC) reactor  \\
                                                            & simulation miniapp  \\
            \texttt{miniSweep}~\cite{hpc5} & A deterministic Sn radiation transport miniapp  \\
            \texttt{SOMA}~\cite{hpc6} & A simulation package for polymeric systems \\ %
            \multirow{2}{*}{\texttt{tealeaf}~\cite{hpc7}} & A miniapp that solves the linear heat conduction  \\
                                                          & equation  \\
            
            \hline
        \end{tabular}
        }
    \end{minipage}

\end{table}

\subsection{Performance Divergence Results}
\label{sec:experiments-overall}

\parh{Microarchitectural Behavior Assessment.}
\T~\ref{tab:experiments-overall} quantifies the microarchitectural
behavior differences for test cases exhibiting significant performance
disparities between compilers.
Column 2 denotes the winning compiler that produces the faster binary.
Columns 3-8 detail the winner's improvement over the other across key
microarchitectural metrics, where negative values signify a higher cost
(e.g., more cache misses).
For instance, Clang version \texttt{adi} executes 5.2\% fewer
instructions than the GCC version but suffers 3.4\% more cache misses.

\begin{table}[!htp]
	\centering
	\caption{Microarchitectural behavior assessment. %
    Columns 3-8 report the winner's improvement over the other, %
    measured by \#instructions executed (\texttt{Ins}), 
    \#CPU micro-ops 
    ($\mu$\texttt{ops}, reflecting instruction complexity), \#memory access
    instructions (\texttt{Mem}), \#cache misses (\texttt{CM}), and \#branch
    mispredictions (\texttt{BM}). Column 9 reports instructions per cycle
    (GCC/Clang). Values with significant differences are marked in bold.} 

    \resizebox{0.80\linewidth}{!}{
		\begin{tabular}{l | c | c  c  c  c  c  c | c}
			\hline 
			\multirow{2}{*}{Program} & \multirow{2}{*}{Win} & \multicolumn{7}{c}{Microarchitectural metrics}\\
            \cline{3-9}
            &  & \texttt{Cycles} & \texttt{Ins} & $\mu$\texttt{ops} & \texttt{Mem} & \texttt{CM} & \texttt{BM} & \texttt{IPC}\\
			
			\hline
			adi    & LLVM  & 13\% & 5.2\% & 6.6\%  & \textbf{17.3\%} & -3.4\% & \textbf{24.8\%} & 0.72/0.79\\
			durbin & LLVM & 5\% & 12.4\% & 5.9\%  & \textbf{-34.7\%} & 5.9\% & \textbf{43\%} & \textbf{1.89/1.74} \\
			gemm   & LLVM & 20.1\% & \textbf{18.6\%} & 12.2\%  & 0\% & 6.7\% & \textbf{49\%} & 2.32/2.39\\
            gesummv	 & LLVM & 5\% & 4.6\% & 0.8\%  & 0.6\% & 1.3\% & -3.5\% & 1.52/1.52 \\
            jacobi-1d	 & LLVM & 8.3\% & \textbf{17.4\%} & 12.6\%  & 10.1\% & -12.2\% & \textbf{39.9\%} & \textbf{3.04/2.74} \\
            jacobi-2d	 & LLVM & 7.1\% & 14\% & 10.4\%  & 8.1\%x & 2.3\% & \textbf{48.9\%} & \textbf{1.93/1.79} \\
            floyd-warshall	 & LLVM & 27.1\% & \textbf{16.7\%} & 10\%  & 0\% & 5.7\% & \textbf{50\%} & \textbf{2.69/3.08} \\
			\hline
            cjpeg	 & GCC & 5.6\% & -1.5\% & -8.9\%  & 2.8\% & 3.7\% & -4.9\% & \textbf{1.82/1.69} \\
            core	 & GCC & 5\%   & 10.1\% & 6.7\%  &-2.6\%  & \textbf{23.2\%} & 14.2\% & \textbf{2.31/2.44} \\
            linpack	 & GCC & 13.3\%&\textbf{-16.8\%} & -10.3\% & 1.5\% & -2\% & \textbf{-69.1\%} & \textbf{3.61/2.68} \\
            loops	 & LLVM & 14.7\%& \textbf{19.7\%} & 9.1\%  & -3.8\% & 7.5\% & \textbf{39.8\%} &\textbf{2.07/1.95} \\
            nnet	 & GCC & 14.4\%& \textbf{18.7\%} & \textbf{16.1\%} & \textbf{17.3\%} & 7.3\% & \textbf{39.1\%} &\textbf{2.18/2.29} \\
            radix2	 & LLVM & 9.3\% &  5.6\% & \textbf{17.5\%} & \textbf{34.5\%} & -0.3\%& \textbf{17.5\%} &\textbf{3.12/3.25} \\
            sha	     & LLVM &  13\% & -0.8\% & -4.0\% & -2.6\% & -4.7\%& -13.6\% &\textbf{2.55/2.96}\\
            zip	     & LLVM & 9.5\% & \textbf{19.6\%} & 14.3\% &  2.1\% & \textbf{27.3\%}& -3.2\% &\textbf{1.95/1.73} \\
            \hline
            perlbench& GCC & 6.0\%& 14.5\% & \textbf{15.5\%} & 14.1\% & 0.9\% & 2.6\% & \textbf{1.65/1.85} \\
            mcf      & GCC & 6.9\%& \textbf{25.3\%} & \textbf{25.2\%} & \textbf{24.9\%} & -0.5\% & 3.4\% & \textbf{0.85/1.05}\\
            x264     & GCC & 15.2\% & \textbf{27.9\%} & \textbf{30.4\%} & \textbf{38.7\%} & 1.7\% & 1.9\% & \textbf{2.62/3.07} \\
            imagick  & LLVM & 12.1\% & \textbf{23.7\%} & \textbf{19.0\%} & \textbf{42.3\%} & \textbf{15.2\%} & \textbf{-22.8\%} & \textbf{3.28/2.83} \\
            nab      & GCC & 38.7\% & 5.8\% & \textbf{26.8\%} & 14.5\% & \textbf{39.3\%} & \textbf{-121.7\%} & \textbf{1.86/1.21}\\
            \hline
            lbm	     & GCC & 11.7\%& 10.3\% & 14.7\% & \textbf{24.3\%} & \textbf{-23.7\%}&11.2\%&1.97/1.94 \\
            flow	 & GCC & 73.6\%& \textbf{79.4\%} & \textbf{81\%} & \textbf{88.7\%} & 9.0\% & \textbf{46.6\%} & \textbf{1.65/2.12}\\
            hot	     & GCC & 80.8\%& \textbf{94.7\%} & \textbf{93.9\%} & \textbf{96.6\%} & \textbf{25\%} & \textbf{86.2\%} & \textbf{0.61/2.23}\\ %
            miniSweep& LLVM &  48\% & \textbf{49.7\%} & \textbf{55.3\%} & \textbf{56.5\%} & \textbf{52.4\%} & \textbf{52.9\%} & \textbf{3.02/2.92}\\ 
            simpleMOC& GCC & 8.9\% & 13.4\% & 11.5\% & \textbf{20\%} & -4.5\% & \textbf{22.3\%} & \textbf{2.31/2.42}\\
            SOMA	 & LLVM & 13.8\%& 13.2\% & \textbf{17.1\%} & \textbf{19.7\%} & 1.7\% & \textbf{18\%} & 2.17/2.18\\
            tealeaf	 & GCC & 31\% & \textbf{65.8\%} & \textbf{66.2\%} & \textbf{64.9\%} & 9.7\% & \textbf{59\%} & \textbf{1.09/2.2} \\
			\hline
		\end{tabular}
	}
	\label{tab:experiments-overall}
\end{table}

The data reveals substantial performance differences, with the \texttt{Cycles}
count improvements ranging from 5\% to 80.8\%.
Our analysis of \T~\ref{tab:experiments-overall} yields several key insights.
First, neither compiler consistently outperforms the other. 
The optimal compiler varies depending on the benchmark (column \texttt{Win}). 
This indicates that each compiler's optimizations have unique strengths and
weaknesses, suggesting significant room for potential improvements in both
compilers.

Second, performance differences stem from compound factors. Consider
\texttt{adi}, LLVM produces a faster binary with reduced
instructions (\texttt{Ins}: 14.2\%), micro-operations (\texttt{$\mu$ops}:
5.2\%), and cache misses (\texttt{CM}: 6.6\%), 
implying advantages in both code efficiency and memory access.
The diverse patterns we observed across metrics indicate a wide range of
potential improvements %
to explore.

Third, a key finding is the strong correlation between performance gains
and instructions per cycle (IPC). For nearly all programs, the compiled
binaries exhibit a significant gap over IPC.
For example, GCC-compiled \texttt{linpack} executes 16.8\% more instructions yet
achieves 13.3\% fewer cycles due to a significantly higher IPC (3.61), indicating
dramatically superior CPU pipeline utilization. 
This underscores our central argument that %
static code properties poorly predict performance, which is largely %
governed by binaries' microarchitectural efficiency.

\begin{table}[!htp]
	\centering
	\caption{Top-3 metric types with most significant differences.} 

    \resizebox{0.817\linewidth}{!}{
		\begin{tabular}{l | c | c | c || l |c | c | c}
			\hline 
			Program & \multicolumn{3}{c||}{Top-3 Metric Types} & Program & \multicolumn{3}{c}{Top-3 Metric Types}\\
			
			\hline
			adi    & ports  & fp\_arith & stlb & jacobi-1d  & memory  &  & \\
            durbin    & ports  & memory & fp\_vector & jacobi-2d  & ports &  &  \\
            gemm    & mite  & ports & fused\_ins & floyd-warshall  & mite & ports & memory\\
            gesummv    & ports & & &  & & \\
            \hline
            \hline
            cjpeg    & fetch  & mite & BM & nnet  & ports & fp\_vector & memory \\
            core    & ports  & mite & L1\_cache & radix2  & ports & fp\_vector & memory\\
            linpack    & mite  & ports & dsb & sha  & ports & L1\_cache &\\
            loops    & ports  & L1\_cache & fp\_arith & zip  & ports &  &\\
			\hline
			\hline
			perlbench    & fetch & light\_ops & memory & imagick & fp\_vector & memory & ports \\
            mcf    & ports  & fetch & dsb & nab  & ports & fp\_arith &
            bad speculation \\
            x264  & ports & memory & light\_ops & & & &  \\
            \hline
            \hline
            lbm    & ports  & divider & fetch & simpleMOC  & L3\_cache & ports & memory\\
            flow    & ports  & memory & stlb & SOMA  & ports & fp\_vector & mite\\
            hot    & ports  & memory & fp\_vector & tealeaf  & ports & fp\_vector & memory\\
            miniSweep    & ports  & memory & mite & & & \\
			\hline
		\end{tabular}
	}
	\label{tab:experiments-metrics}
\end{table}

\parh{Drilling Down by Metric Types.}
\T~\ref{tab:experiments-metrics} ranks the top-3 microarchitectural metric
types with the most significant %
differences for each benchmark. 
We identified a variety of critical
metrics that not only helped us understand the root causes %
from multiple perspectives but also confirmed our suspicion that performance
differences are affected by multiple microarchitectural factors that have long
been overlooked by prior work. 
The hierarchical analysis pinpoints the critical hardware behaviors, 
highlighting four main categories of
performance issues.

\parhs{1) Low IPC and Execution Port Pressure.}
This is the most common divergence type we identified.
Ports in modern CPUs act as interfaces, connecting the instruction scheduler to
the various execution units that perform computations~\cite{abel2019uops}. Each
port is typically associated with specific types of operations or execution
units (e.g., arithmetic logic units, load/store units, or specialized units for
tasks like division). 
Differences in such metrics typically indicate contention for execution units.
This suggests divergent compiler optimizations for instruction-level
parallelism, often due to suboptimal scheduling or long data
dependencies~\cite{wall1991limits}.

\parhs{2) Memory-Related Issues.}~Performance disparities caused by issues like
L1/L3 cache misses and second-level data TLB (STLB) misses are prevalent in
data-intensive benchmarks. Significant differences in these metrics reveal
defective compiler optimizations in reducing redundant memory accesses, managing
data locality, and optimizing cache usage.

\parhs{3) Floating Point (FP) and Vectorization.}~Differences in FP operations
and vectorization optimizations reflect variations in the compilers' ability to
effectively utilize CPU vector units, which are crucial for performance in
affected benchmarks.

\parhs{4) Frontend and Instruction Supply Bottlenecks.}~%
Several benchmarks show significant divergence in frontend metrics, which relate
to the CPU's ability to fetch and decode instructions to feed the execution
backend efficiently. 
Modern processor frontend designs feature multiple paths that micro-operations
can take~\cite{xu2022reverse,deng2022processor}: through the Micro-Instruction
Translation Engine (MITE), the Decode Stream Buffer (DSB),
or the Loop Stream Detector (LSD). Each path shows unique timing
and power signatures. 
Inefficient use of frontend paths can cause instruction fetch
latencies~\cite{deng2022leaky} that starve CPU execution units, leading to
pipeline stalls despite available backend resources.

\smallskip
\begin{tcolorbox}[size=small,colback=gray!5!white,colframe=gray!75!black,]
\parh{Finding 3 (RQ2):}~The performance gaps (``unseen deltas'') introduced by
contemporary compiler optimizations are both substantial and widespread.
They are fundamentally rooted in differences in microarchitectural behaviors,
specifically in execution port utilization, memory access patterns,
vectorization, and instruction fetch.
\tool\ successfully identifies these critical yet overlooked drivers of
performance differences by emphasizing the top-down exploration of the
hierarchical microarchitectural metrics.
\end{tcolorbox}

\subsection{Root Cause Analysis}
\label{sec:analysis}

Having identified critical performance deviations through microarchitectural
metrics, we examined the underlying low-level code to determine
their root causes. Our microarchitecture-aware manual investigation 
reveals eight primary root-cause patterns, corresponding to the
performance issues identified in \S~\ref{sec:experiments-overall}. For each
pattern, we present concrete examples that illustrate how subtle machine-code
divergences can lead to significant performance gaps.

\begin{figure}[h]
    \centering
    \includegraphics[width=0.83\linewidth]{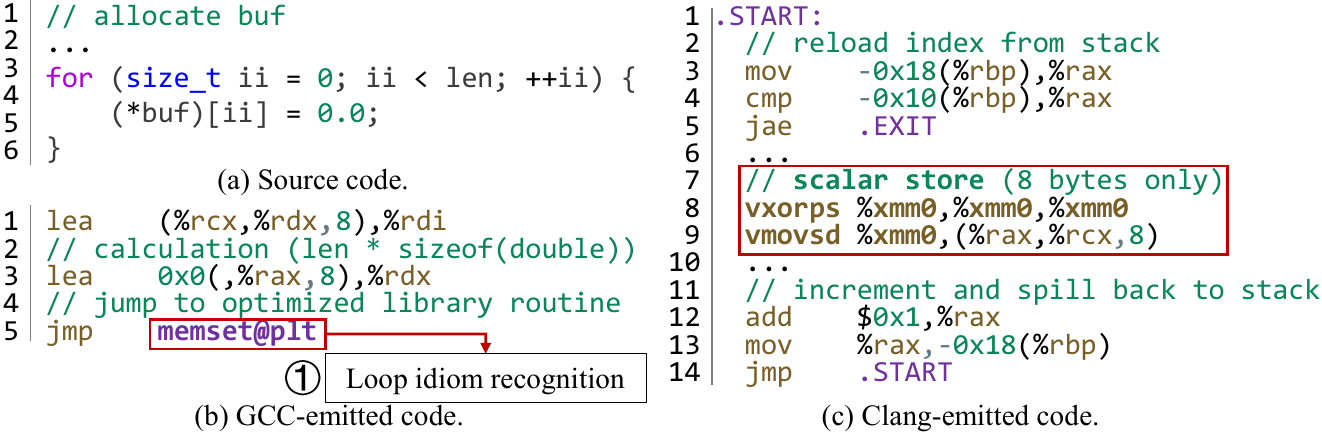}

    \caption{Divergence for array initialization in
    in the \texttt{flow} benchmark: Scalar stores (Clang) vs. \texttt{memset}
    (GCC).}
    \label{fig:case1}
\end{figure}

\parhs{1) Inefficient Memory Access Patterns.}~Suboptimal memory access, often
missed by static analysis but evident in microarchitectural metrics, %
can severely degrade performance.
Compilers may introduce such inefficiencies by generating redundant memory
operations
or accessing data less efficiently. 
Our analysis of the \texttt{flow} benchmark revealed a significant
divergence in memory-related metrics. \F~\ref{fig:case1} illustrates a primary
source of this inefficiency. The source code (\F~\mysubref{fig:case1}{a})
contains a \texttt{for} loop that zero-initializes an array of double-precision
floating-point numbers. While Clang-emitted code adheres strictly to the loop's
syntactic structure by iteratively writing 8-byte memory with scalar stores, GCC
identifies that the floating-point value \texttt{0.0} corresponds to an all-zero
bit pattern. Consequently, it utilizes loop idiom recognition to replace the
explicit loop with a call to %
\texttt{memset}
(\circled{1}), whose implementation is highly tuned with wide SIMD registers,
allowing it to effectively saturate the memory bandwidth.

\begin{figure}[h]
    \centering
    \includegraphics[width=0.999\linewidth]{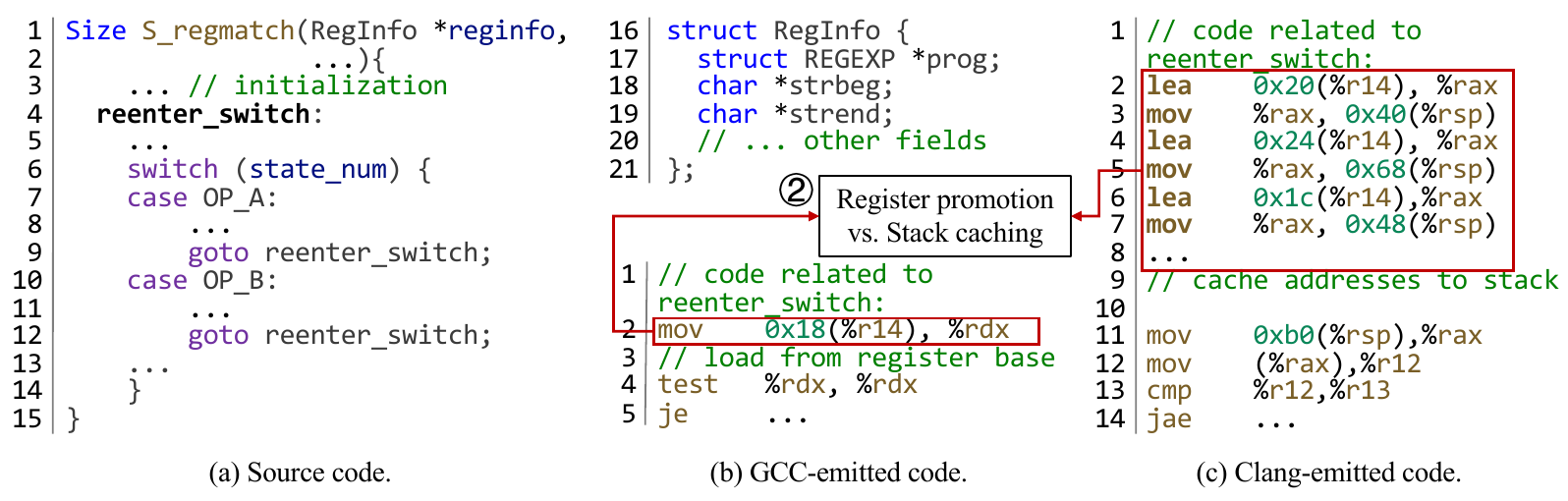}

    \caption{Register allocation divergence in the \texttt{perlbench}
    benchmark. GCC optimized context access via registers while Clang emitted a
    costly sequence of stack stores.}
    \label{fig:case1-spec}
\end{figure}

We also observed a significant divergence in how compilers handle
structure member access within a loop exhibiting high register pressure,
specifically within the regular expression engine (\texttt{S\_regmatch}) of the
\texttt{perlbench} benchmark (\F~\ref{fig:case1-spec}). As illustrated in
\F~\mysubref{fig:case1-spec}{a}, the function contains a large \texttt{switch}
statement that creates intense demand on physical registers.
Clang adopts a conservative strategy to ensure consistent register availability
for the loop's worst-case complexity. It inserts a block of \texttt{lea} and
\texttt{mov} instructions at the jump target to pre-calculate structure member
addresses and proactively spill them to the stack. While safe, this strategy
incurs a high penalty in both instruction count and memory accesses per
iteration.
Conversely, GCC optimizes for the average case by assigning critical variables
to callee-saved registers (e.g.,
\texttt{r14}). This strategy avoids the per-iteration preamble overhead by
performing arithmetic offsets lazily only when data is required, assuming
sufficient register capacity is available without aggressive spilling.

\begin{figure}[h]
    \centering
    \includegraphics[width=0.999\linewidth]{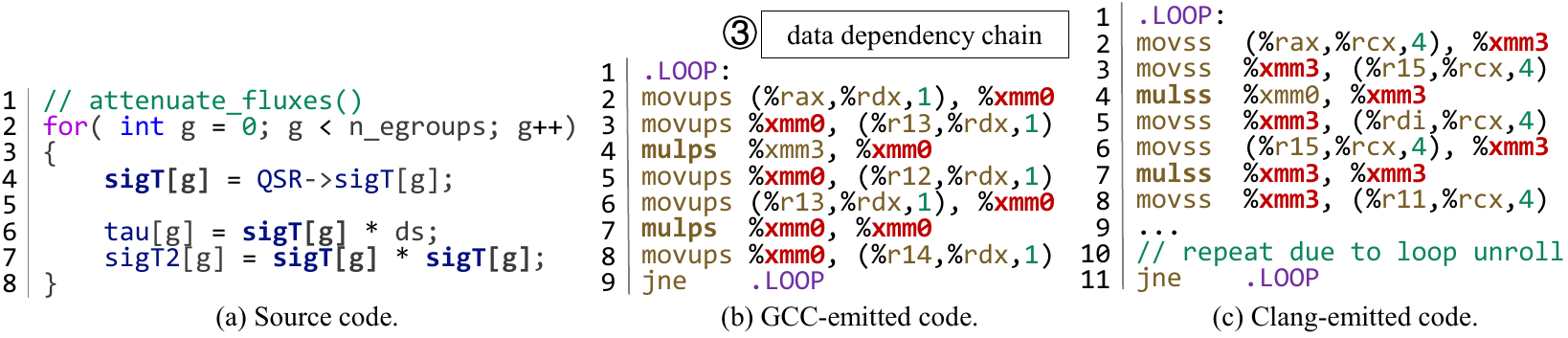}

    \caption{Suboptimal instruction-level parallelism in the \texttt{simpleMOC}
    benchmark.}
    \label{fig:case2}
\end{figure}

\parhs{2) Suboptimal Instruction-Level Parallelism.}~Modern superscalar
processors require a steady stream of independent instructions to maximize
instruction-level parallelism (ILP). 
A significant barrier to ILP arises when low-level code forms long
data dependency chains, forcing the processor to stall while execution
units remain idle.
In the \texttt{simpleMOC} benchmark, our inspection revealed that the
Clang-generated code suffered from severe microarchitectural stalls and low
parallelism (\F~\mysubref{fig:case2}{c}).
Despite applying loop unroll, Clang creates a tight dependency
chain centered on \texttt{xmm3}, hindering out-of-order execution. 
Furthermore, Clang floods the execution ports with scalar micro-ops
(\texttt{mulss}), hitting the throughput limit of decode and dispatch engines.
In contrast, while GCC-emitted code also suffers from similar data-dependency
issues, GCC uses execution ports more efficiently by utilizing packed
instructions (\texttt{mulps}). By processing four floating-point values per
instruction, 
GCC amortizes the latency of the dependency chain and achieves higher throughput
than Clang's scalar implementation.

\begin{figure}[h]
    \centering
    \includegraphics[width=0.86\linewidth]{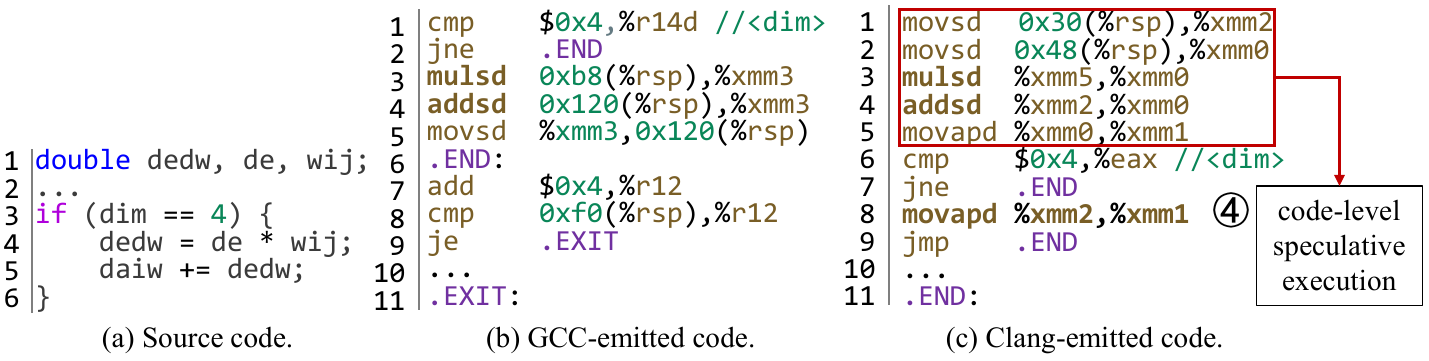}

    \caption{Performance degradation due to unsafe speculative execution in the \texttt{nab}
    benchmark.}
    \label{fig:case2-spec}
\end{figure}

Besides, aggressive compiler optimizations designed to mitigate
control hazards can backfire when they expose hardware to pathological data
patterns. Techniques such as if-conversion and instruction hoisting aim to
eliminate branch misprediction penalties by converting control dependencies into
data dependencies; however, this strategy can force the execution of operations
that were originally guarded by conditional logic.
In the example illustrated in \F~\ref{fig:case2-spec}, Clang hoists the
double-precision multiplication (\texttt{mulsd}) above the branch check to
execute it speculatively. While this theoretically improves pipeline
utilization, it indiscriminately processes input data even when the condition is
false. In these invalid iterations, the operations often involve denormalized
numbers (extremely tiny numbers that fall below the range of normal
floating-point representation, e.g., $2\times10^{-308}$) due to uninitialized
garbage data.
Since modern FPUs are optimized for normalized arithmetic, processing denormals
typically triggers a floating point assist. This forces the processor to flush
the pipeline 
and incur a penalty of hundreds of cycles, severely degrading performance.

\parhs{3) Instruction Fetching and Frontend Stalls.}~The processor's 
frontend fetches, decodes, and supplies instructions to the
execution backend. Bottlenecks here can starve the backend, causing pipeline
stalls even when execution resources are available. These issues are often
caused by subtle code layout and instruction choices. %

\begin{figure}[ht]
    \centering
    \begin{minipage}{0.51\textwidth}
        \centering
        \includegraphics[width=0.81\linewidth]{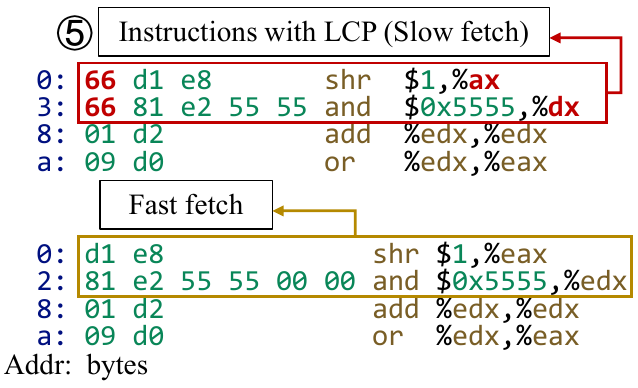}

        \caption{Frontend stalls caused by length-changing prefix in the
        \texttt{core} benchmark compiled with GCC (above). }
        \label{fig:case3}
    \end{minipage}
    \hfill
    \begin{minipage}{0.48\textwidth}
        \centering
        \includegraphics[width=0.91\linewidth]{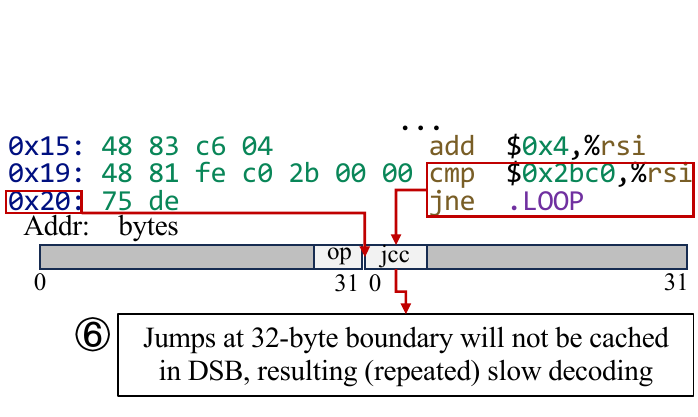}

        \caption{Decode stream buffer inefficiency in \texttt{floyd-warshall}
        compiled with GCC.}
        \label{fig:case4}
    \end{minipage}
\end{figure}

Our analysis of the \texttt{core} benchmark revealed a high rate of frontend
stalls in the GCC binary.
\tool\ traced this inefficiency to the critical code shown in
\F~\ref{fig:case3}. 
Specifically, GCC generates the \texttt{and} and \texttt{shr} instructions with
length-changing prefixes (LCP)~\cite{deng2022leaky}.
These prefixes alter the length of the instruction's opcode and operand field,
specifically affecting the instruction's byte length.
Here, an immediate operand (\texttt{0x5555}) is changed from 32-bit
to 16-bit (\circled{5}).
At the microarchitectural level, LCPs can cause delays in instruction decoders,
which are optimized for predictable instruction boundaries~\cite{deng2022leaky,abel2023facile}.
This delay can create a ``bubble'' in the pipeline that slows the instruction fetch rate.
An optimized ``fast fetch'' version avoids this by using
alternative instructions without LCPs, ensuring a smooth supply of instructions.

Similarly, in the \texttt{floyd-warshall} benchmark, \tool\ identified high
frontend instruction decoding costs in the GCC binary due to a subtle code layout
decision.
As demonstrated in \F~\ref{fig:case4}, 
GCC places a conditional jump
(\texttt{jne}) precisely across a 32-byte boundary (\circled{6}). 
Intel processors utilize the DSB to cache decoded micro-operations for
small loops, bypassing the complex and slower legacy decoders (MITE).
However, the DSB cannot cache jumps that cross a 32-byte boundary~\cite{intel-jcc-erratum}. 
This suboptimal alignment prevents the DSB from caching the jump, forcing the
CPU to re-decode the instruction on every iteration and causing a significant,
persistent performance penalty.

\begin{figure}[h]
    \centering
    \includegraphics[width=0.95\linewidth]{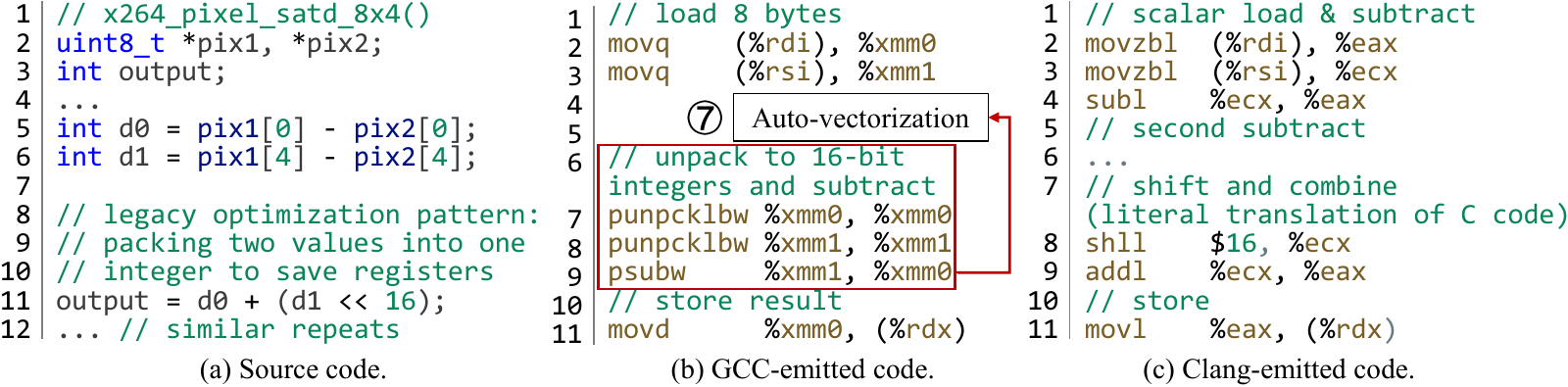}

    \caption{GCC's auto-vectorization resilience to legacy optimization patterns
in the \texttt{x264} benchmark.}
    \label{fig:case5-spec}

\end{figure}

\parhs{4) Missed Vectorization and Underutilization of SIMD Units.}~The use of
Single Instruction Multiple Data (SIMD) instructions is critical for maximizing
throughput in computationally intensive tasks. A compiler's failure to
effectively vectorize code represents a high-impact optimization defect.
We traced the performance divergence in the pixel analysis kernel of
the \texttt{x264} benchmark (\F~\ref{fig:case5-spec}) to Clang's handling of
manual bitwise manipulation. The source code employs manual packing logic to
combine data elements.
Clang performs a literal translation of the source, executing scalar arithmetic
on general-purpose registers. This serialized approach forces the CPU to process
pixels individually, emulating the packing logic via ALU operations rather than
exploiting the hardware's vector capabilities. Furthermore, the reliance on
scalar registers induces severe register pressure, leading to excessive stack
memory spills.
Conversely, the GCC successfully identifies the underlying parallelism, lowering
the manual bit manipulation into native packed vector instructions (e.g.,
\texttt{psubw}). This enables simultaneous execution on multiple data points
within 128-bit XMM registers and drastically reduces memory traffic latency.

\begin{figure}[h]
    \centering
    \includegraphics[width=0.92\linewidth]{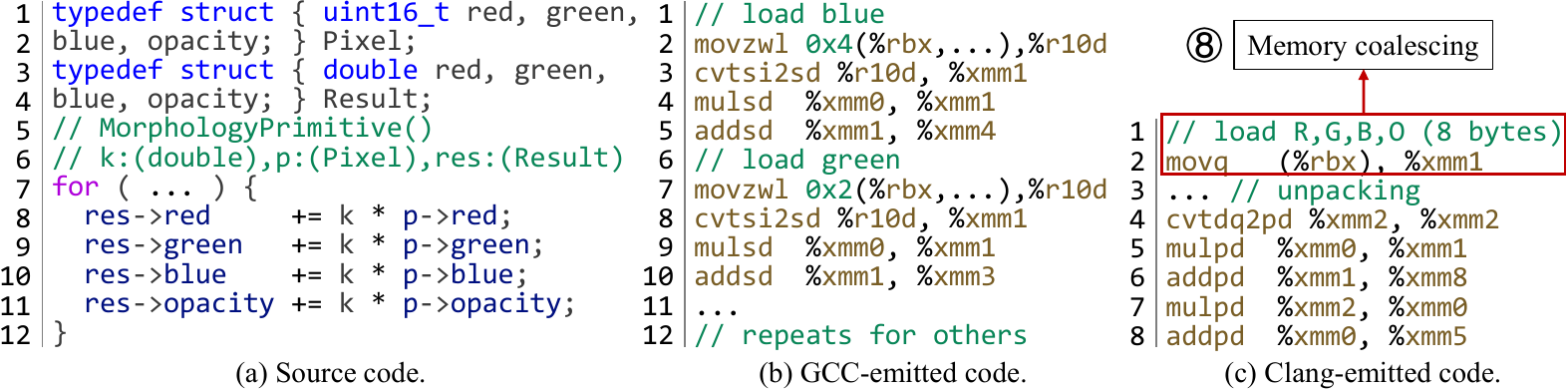}

    \caption{Clang's utilization of Superword-Level Parallelism (SLP)
    for memory coalescing versus GCC's scalar expansion in the
    \texttt{imagick} benchmark.}
    \label{fig:case6-spec}

\end{figure}

Additionally, access patterns over contiguous structured data offer
significant opportunities for Superword-Level Parallelism (SLP), where adjacent
memory loads and independent operations can be fused into vector instructions. A
compiler's inability to identify these patterns leads to redundant memory
traffic and pipeline underutilization. 
This discrepancy is evident in the \texttt{iamgick} benchmark, where GCC fails
to vectorize the pixel accumulation loop shown in \F~\ref{fig:case6-spec}.
While GCC-emitted code 
issues four separate load operations per iteration and utilizes scalar
arithmetic, leaving the upper lanes of the SIMD registers idle, Clang
successfully exploits the data layout by coalescing the four 16-bit loads into a
single 64-bit memory access. Furthermore, Clang promotes the scalar operations
to packed double-precision arithmetic (\texttt{mulpd}, \texttt{addpd}),
processing color channels in parallel. This transformation effectively doubles
the computational throughput and reduces memory instruction overhead,
significantly optimizing the processor's execution pipeline.

\smallskip
\begin{tcolorbox}[size=small,colback=gray!5!white,colframe=gray!75!black,]
\parh{Finding 4 (RQ3):}~The root causes of performance differences often stem from
subtle, low-level code optimization decisions invisible to
analysis based on static code properties.
Our top-down differential analysis effectively identifies these critical issues
and pinpoints specific code snippets for detailed investigation.
The %
root causes highlight systemic challenges in compiler
optimizations 
and underscore the need for more comprehensive testing and
investigation %
that accounts for the complex code-microarchitecture interplay.
\end{tcolorbox}

This work provides a powerful new lens for understanding the complex
interactions between compiler optimizations and microarchitectural behavior and
offers insights for improving compiler optimizations.
Moreover, the approach is far from saturated; when combined with existing source
code transformation
techniques~\cite{sun2016finding,li2024boosting,zhang2017skeletal,wu2025compiler,gao2024shoot},
we anticipate it will uncover a wide and diverse range of compiler optimization
defects and performance-harming code patterns.

\subsection{Post-Compilation Binary Patching}
\label{sec:patching}

To quantitatively assess the impact of issues we identified, we
developed a post-compilation binary patching framework, which allows us
to modify a slower binary's machine code directly, replacing suboptimal
code snippets with more efficient versions generated by a different compiler. 
This process provides concrete evidence that the microarchitectural behavior
differences identified in \S~\ref{sec:analysis} are indeed the root cause of
observed performance gaps. 
By rectifying these localized inefficiencies, we achieve
substantial whole-program improvements, offering powerful validation for our
top-down differential analysis.

\begin{wrapfigure}[10]{l}{0.62\linewidth}
    \centering
    \includegraphics[width=0.999\linewidth]{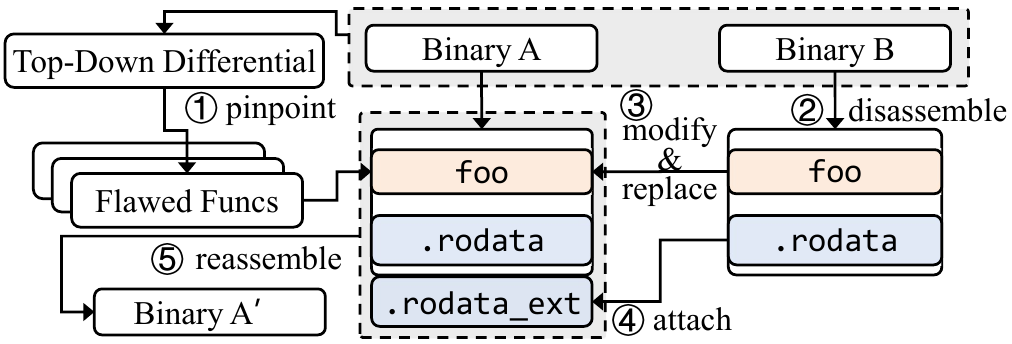}

    \caption{Illustration of the binary patching workflow. }
    \label{fig:patching}
\end{wrapfigure}

Our systematic binary patching workflow, illustrated in \F~\ref{fig:patching},
integrates superior code from a ``donor'' binary ($B$) into a ``recipient''
binary ($A$). The process begins by using the results of top-down
differential analysis to pinpoint the specific flawed functions within $A$ that
are responsible for the performance degradation (\circled{1}), while
simultaneously identifying their counterparts in $B$.  
Next, both binaries are disassembled using advanced reassembleable disassembling
techniques to ensure the compilability of disassembled
code~\cite{wang2015reassembleable,dinesh2020retrowrite} (\circled{2}).
The flawed function in $A$ (e.g., \texttt{foo}) is then replaced with the
efficient code from $B$ (\circled{3}).\footnote{We limit patching to
functions without calls due to the complexities inherent in handling function
calls and pointers. Our goal is not to guarantee perfect patching for every
function. Instead, our primary objective is to demonstrate the significant
performance impact of identified inefficiencies and to offer promising, albeit
preliminary, evidence that our findings can effectively guide (binary-level)
optimization enhancement.}
This ``transplant'' may also require modifying and attaching the read-only data
section (\texttt{.rodata}) if the new code relies on different constants or data
structures (\circled{4}). 
Finally, the modified assembly code is reassembled into a new functional
executable ($A'$) for performance comparison with the original $A$.

\begin{table}[!htp]
    \centering

    \begin{minipage}{0.60\textwidth}
        \centering
        \caption{Performance improvements after patching. (``\texttt{NA}'': no
        patchable functions or no observable improvement.)}

        \resizebox{0.93\linewidth}{!}{
            \begin{tabular}{l | c | c || l |c | c}
                \hline 
                \multirow{2}{*}{Program} & \multicolumn{2}{c||}{Improvement} & \multirow{2}{*}{Program} & \multicolumn{2}{c}{Improvement}\\
                & \multicolumn{1}{c}{GCC} & \multicolumn{1}{c||}{Clang} &   & \multicolumn{1}{c}{GCC} & \multicolumn{1}{c}{Clang} \\
                
                \hline
                adi    & 13.3\%  & NA  & jacobi-1d  & \textbf{11.5\%}  & NA  \\
                durbin    & 5.9\%  & NA  & jacobi-2d  & 6.5\% & NA    \\
                gemm    & 21.4\%  & \textbf{6.3\%} & floyd-warshall  & 14.4\% &  NA\\
                gesummv    & 6.4\% & NA & &  \\
                \hline
                \hline
                cjpeg    & NA  & 6.6\% & nnet  & NA & 14.7\%  \\
                core    & \textbf{8.7\%}  & NA & radix2  & \textbf{12.4\%} & 1.2\% \\
                linpack    & \textbf{10.2\%}  & 19.7\% & sha  & 12.7\% & NA \\
                loops    & 12.5\%  & \textbf{3.2\%} & zip  & \textbf{19.3\%} & 6.4\% \\
                \hline
                \hline
                perlbench    & NA  & NA & imagick  & NA & NA  \\
                mcf    & NA  & 1.4\% & nab & NA & NA \\
                x264  & \textbf{2.0\%} & \textbf{16.1\%} & & &  \\
                \hline
                \hline
                lbm    & \textbf{5.2\%}  & 16.2\% & simpleMOC  & NA & 1.4\% \\
                flow    & NA  & 68.1\% & SOMA  & 6.0\% & NA \\
                hot    & NA  & 80.5\% & tealeaf  & NA & 30.8\% \\
                miniSweep    & 48.6\%  & NA  & & &  \\
                \hline
            \end{tabular}
        }
        \label{tab:patching-data}
    \end{minipage}
    \hfill %
	\begin{minipage}{0.37\textwidth}
        \centering
        \caption{Performance of patched binaries compared to the best compiler-generated
        versions.} 

        \resizebox{0.68\linewidth}{!}{
            \begin{tabular}{l | c }
                \hline 
                Program & Improvement \\
                
                \hline
                gemm    & 6.3\%     \\
                core    & 8.7\%   \\
                linpack    & 10.2\%    \\
                loops    & 3.2\%   \\
                
                x264 & 2.0\%  \\
                radix2  & 3.1\% \\
                zip  & 10.8\% \\
                lbm  & 5.2\% \\
                \hline
            \end{tabular}
        }
        \label{tab:patching-winner}
    \end{minipage}

\end{table}

The results, detailed in \T~\ref{tab:patching-data}, demonstrate significant
performance improvements across most benchmarks. The only exceptions
are a few complex cases, marked with \texttt{NA} for both compilers, where
identified functions cannot be easily patched. Where patching was feasible, the
gains were substantial. Specifically, patching Clang-compiled \texttt{hot} and
\texttt{flow} benchmarks
yielded performance gains of 80.5\% and
68.1\%, respectively. 
Similarly, patching GCC-generated \texttt{gemm},
\texttt{miniSweep}, and \texttt{linpack} yielded improvements of 21.4\%, 48.6\%,
and 10.2\%. 
These results confirm that the suboptimal code patterns identified in
\S~\ref{sec:analysis} were indeed the cause of performance gaps reported in
\T~\ref{tab:experiments-overall}.

Furthermore, \T~\ref{tab:patching-winner} compares the patched binaries against
the best-performing binaries generated by any compiler.
Remarkably, for 8 out of 27 benchmarks, patched binaries outperformed the
compiler-generated winners, with improvements of up to 10.8\%. 
This demonstrates the practical applicability of our approach, suggesting that
cross-compiler insights can be leveraged to mitigate performance degradation and
guide future compiler optimizations.

\smallskip
\begin{tcolorbox}[size=small,colback=gray!5!white,colframe=gray!75!black,]
\parh{Finding 5 (RQ3):}~The dramatic speedups achieved by replacing just a few
critical functions validate our focus on microarchitectural behaviors to uncover
the most impactful performance issues and confirm that the root causes we
identified are indeed responsible for the observed performance gaps.
Moreover, by identifying and transplanting superior code, we not only validate
the root causes of performance differences but also showcase a tangible method
for rectifying them, paving the way for more flexible and targeted binary-level
performance improvements.
\end{tcolorbox}

\section{Discussion}
\label{sec:discussion}

\parh{Generalizability across CPUs and PMUs.} Although our evaluation is
conducted on an Intel CPU, the \tool methodology itself is not tied to any
particular processor. %
The top-down hierarchy it builds on is a vendor-independent model with
established counterparts on AMD, ARM, and RISC-V
processors~\cite{yasin2014top,jarus2016top,mou2024top,arm-mca}.
Hardware-assisted sampling is likewise available across these platforms, and the
differential comparison and debug-information-based code mapping are
hardware-agnostic. Porting \tool to a new platform therefore primarily requires
re-deriving its event set, %
since event names and hierarchy granularity vary across vendors. As for the
findings, most root causes identified in \S~\ref{sec:analysis} (e.g.,
inefficient memory access, suboptimal register allocation, limited
instruction-level parallelism, missed vectorization or superword-level
parallelism) stem from properties fundamental to any out-of-order superscalar
design and are thus expected to transfer broadly, although exact cycle penalties
will differ. Two patterns, length-changing-prefix stalls and
decode-stream-buffer alignment, are specific to Intel's frontend and may not
reproduce identically on other microarchitectures. We regard this
microarchitecture sensitivity as intrinsic to performance analysis rather than a
limitation of our approach. A severe stall on a widely deployed Intel CPUs
remains a legitimate, high-impact defect for that ecosystem, regardless of
whether other processors happen to mask it.

\section{Related Work}
\label{sec:related}

A significant body of research has focused on compiler correctness, primarily by
generating diverse test programs to uncover functional bugs like crashes and
miscompilations. 
For instance, Hermes finds bugs by actively mutating executed
code~\cite{sun2016finding}.
Creal incorporates test
programs with code fragments from real-world projects~\cite{li2024boosting},
while another recent study uses actual bug reports to generate targeted test
cases likely to expose compiler flaws~\cite{zhong2022enriching}.
Other methods prioritize systematic program generation. Zhang et al.,
for example, propose a rigorous testing method by exhaustively generating all
possible programs that fit a given structural template~\cite{zhang2017skeletal}. 
To specifically validate optimizations, Wu et al. apply source-level
equivalence transformations to test whether a compiler's optimizations preserve
program semantics~\cite{wu2025compiler}.

A growing research direction explores the vast space of compilation options.
COTest systematically tests combinations of optimization flags to uncover bugs
triggered by specific compilation sequences~\cite{chen2022boosting}. 
\textsc{Atlas} refines this with fine-grained, source-level attributes to
control optimizations per function or loop~\cite{wu2025unveiling}. 
One recent work
validates JIT compilers by exploring their unique runtime compilation
choices~\cite{li2025validating}. 
Similarly, \textsc{Optimuzz} uses fuzzing targeted at optimization passes with
continuous validation to detect miscompilations~\cite{kwon2025optimization}.
While these works are highly effective for functional correctness, our work
addresses the orthogonal challenge of performance bugs.

\section{Conclusion}

This paper introduces a novel top-down methodology for identifying and analyzing
compiler optimization defects. Our approach is microarchitecture-aware, tracing
end-to-end performance deltas between binaries back to their root causes. 
Through a comprehensive empirical study, 
we have demonstrated that these performance issues %
are both widespread and
substantial in real-world applications. 
This work provides a powerful new lens for understanding the complex interplay
between compiler optimizations and modern hardware, paving the way for more
effective techniques for compiler optimization testing and performance
improvement.

\section*{Data Availability}

To support open science and facilitate the reproducibility of our results, we
have made the research artifact for \tool\ %
available at ~\cite{artifact}. 

\section*{Acknowledgements}

This paper was supported in part by the Fundamental and Interdisciplinary
Disciplines Breakthrough Plan of the Ministry of Education of China (No.
JYB2025XDXM118), the “111 Center” (No. B26023), a grant from the Research
Grants Council of the Hong Kong Special Administrative Region, China HKUST (No.
R6005-25), a RGC GRF grant under the contract 16214723, and an ITF grant under
the contract ITS/161/24FP. Zhibo Liu was supported by the RGC Post-doctoral
Fellowship Scheme (No. PDFS2324-6S08). The authors would like to thank the
Collaborative Innovation Center of Novel Software Technology and
Industrialization, Jiangsu, China, for its support.

\bibliographystyle{ACM-Reference-Format}
\bibliography{bib/compiler, bib/testing, bib/performance}

\end{document}